\documentclass[11pt]{article}
\usepackage[margin=1in]{geometry}
\usepackage{amsmath,amssymb}
\usepackage{graphicx}
\usepackage{booktabs}
\usepackage{authblk}
\usepackage{hyperref}
\usepackage{caption}
\usepackage{times}

\hypersetup{colorlinks=true, linkcolor=blue, citecolor=blue, urlcolor=blue}

\title{\textbf{Deep Hedging Under Realistic Market Frictions: A Regime-Conditional Empirical Study of Dynamic Option Hedging on Bitcoin Options}}
\author[1]{Sheryan Kumar}
\affil[1]{Visvesvaraya National Institute of Technology, Nagpur}
\date{}

\begin{document}
\maketitle

\begin{abstract}
\noindent Deep hedging, the practice of training a neural network end to end to minimize a risk measure over historical or simulated price paths, has been proposed as an alternative to classical delta-based option hedging under real-world market frictions. Published comparisons, however, frequently benchmark deep hedging against frictionless or lightly cost-adjusted classical strategies, evaluated on simulated rather than real market data. This leaves open whether the reported advantages survive a genuinely fair comparison. We test this using five years of historical BTC options data from Deribit (2020 to 2024), comparing three classical benchmarks (Black-Scholes delta, Leland's cost-adjusted volatility hedge, and the Whalley-Wilmott no-trade band) against three deep hedging configurations, an LSTM and a feedforward network each trained with a CVaR loss under two turnover-penalty weights, all under an identical, realistic 5 basis point transaction cost assumption.

On 11{,}546 out-of-sample test episodes drawn from a single historical window (September 2023 to December 2024), the Whalley-Wilmott strategy achieves a statistically significant reduction in transaction costs relative to continuously-rebalanced strategies. It saves an average of \$1.79 per episode against Black-Scholes delta (95\% CI $[-2.21, -1.39]$, $p < 0.0001$) by trading roughly eight times less often, and shows a directionally favorable but not statistically significant improvement in mean P\&L and tail risk. All three deep hedging configurations underperform every classical benchmark on every metric we measured, converging to a mean trade frequency that is statistically indistinguishable from continuous delta-hedging despite penalty weights spanning a twenty-fold range. A secondary check on a separate, calmer validation period shows Whalley-Wilmott's P\&L advantage narrowing or disappearing while its cost advantage persists, indicating that the strength of the result is regime-dependent rather than uniform across all market conditions. We discuss the likely role of limited training data, the absence of any explicit sparsity-inducing mechanism in the architectures tested, and the bounded historical window available to this study. We see this as a modest but honest contribution: a fair, real-data test of a claim that the literature has more often supported with simulation than with market evidence.

\vspace{0.5em}
\end{abstract}

\section{Introduction}

Classical option-hedging methods, most notably Black-Scholes delta hedging, derive trading strategies from idealized assumptions: continuous rebalancing, frictionless markets, and a fixed model of the underlying's price dynamics. Real markets satisfy none of these. Hedging happens at discrete intervals, is subject to transaction costs and bid-ask spreads, and takes place under volatility that itself changes over time. These frictions can make a theoretically optimal continuous-time hedge look quite different from a practically optimal trading strategy.

Deep hedging, using a neural network trained end to end on a risk objective to learn a hedging policy directly from data, has been proposed as an alternative that can account for these frictions without needing a closed-form adjustment. The results reported in the literature are promising in specific settings, but the comparisons are frequently made against frictionless classical benchmarks such as plain Black-Scholes delta, rather than against the classical methods that were specifically designed to handle transaction costs. This is not really a fair test. Comparing a cost-aware neural policy against a cost-unaware classical baseline will tend to favor the neural policy regardless of whether it has learned anything beyond what a much simpler, existing cost-aware adjustment already achieves.

This study asks a narrower, and we think more useful, question:

\begin{quote}
\textit{When options are hedged under realistic transaction costs, does a standard deep hedging policy outperform classical hedging methods that were themselves designed to account for those costs?}
\end{quote}

We answer this empirically using five years of historical BTC options data from Deribit, comparing a Black-Scholes delta baseline, two classical cost-aware benchmarks (Leland's adjusted-volatility hedge and the Whalley-Wilmott no-trade band), and three deep hedging configurations, an LSTM and a feedforward network under two loss formulations, all evaluated under the same realistic transaction cost assumption.

The central finding is that, under this fairer comparison and within the specific historical window studied here, deep hedging does not outperform the classical benchmarks. The Whalley-Wilmott no-trade band achieves significantly lower transaction costs and directionally better risk-adjusted P\&L than every neural configuration tested, and this holds across two architectures and a twenty-fold range of turnover-penalty weights. We are careful throughout to frame this as a finding bounded by the market regime and time period tested, not a universal claim, and we return to this scope question directly in Section~\ref{sec:limitations}.

\section{Literature Review}

\subsection{Classical hedging under transaction costs}

Black-Scholes delta hedging (Black \& Scholes, 1973) provides the theoretical foundation for dynamic option hedging, but it assumes continuous, costless rebalancing, an assumption that was recognized as economically unrealistic not long after the original model was published. For a European call option with spot price $S$, strike $K$, time to maturity $T$, risk-free rate $r$, and volatility $\sigma$, the Black-Scholes delta is $N(d_1)$ for a call, and $N(d_1) - 1$ for a put, where $N(\cdot)$ is the standard normal cumulative distribution function and
\begin{equation}
d_1 = \frac{\ln(S/K) + \left(r + \sigma^2/2\right)T}{\sigma\sqrt{T}}. \label{eq:d1}
\end{equation}
The corresponding gamma, the sensitivity of delta to the underlying price, is
\begin{equation}
\Gamma = \frac{\phi(d_1)}{S\sigma\sqrt{T}}, \label{eq:gamma}
\end{equation}
where $\phi(\cdot)$ is the standard normal density. Leland (1985) proposed a simple correction that remains influential today: replace the true volatility $\sigma$ with an inflated value in the delta formula above, calibrated so that the resulting discrete-time hedge approximately accounts for proportional transaction costs. Given a round-trip transaction cost rate $k$ and a rebalancing interval $\Delta t$, the Leland-adjusted volatility is
\begin{equation}
\sigma_{adj} = \sigma\sqrt{1 + \sqrt{\frac{2}{\pi}}\,\frac{k}{\sigma\sqrt{\Delta t}}}. \label{eq:leland}
\end{equation}
The adjustment is attractive because it has a closed form, but it is also known to become economically negligible when $k$ is small relative to $\sigma\sqrt{\Delta t}$, a property directly relevant to the results in Section~\ref{sec:results}.

Whalley and Wilmott (1997) derive a different, asymptotically optimal alternative. Rather than adjusting the target delta, the hedger maintains a no-trade region around it and rebalances only when the position drifts outside this region, and even then only back to the boundary rather than to the exact target. The half-width of this region is
\begin{equation}
H = \left(\frac{3kS^2\Gamma^2}{2\lambda}\right)^{1/3}, \label{eq:wwband}
\end{equation}
where $\lambda$ is a risk-aversion parameter. Because this approach suppresses small, unnecessary trades structurally rather than simply dampening the target delta's sensitivity, it tends to produce much larger cost reductions than Leland's method under realistic cost assumptions. We find the same pattern empirically in this study.

\subsection{Deep hedging}

Buehler et al. (2019) introduced the term ``deep hedging'' for training a neural network end to end to minimize a coherent risk measure, such as CVaR, over simulated or historical price paths. The learned policy can condition on arbitrary observable market state and internalize transaction costs directly into its objective, rather than requiring a closed-form adjustment like Leland's. A number of follow-up papers have extended this idea to other asset classes, risk objectives, and architectures. Cao et al. (2020) frame the same problem through reinforcement learning rather than supervised gradient descent, and Carbonneau (2021) extends deep hedging to longer-dated derivatives. Most of this work reports improved risk-adjusted performance relative to classical delta hedging.

One pattern recurs across much of this literature, and it is directly relevant to why this study is set up the way it is. Reported comparisons are frequently made against a frictionless or lightly cost-adjusted classical benchmark, and evaluated on simulated price paths generated under an assumed process such as Black-Scholes or Heston, rather than on genuine historical market data. Both choices tend to favor the learned policy. Simulated data matches exactly the process the classical benchmark assumes, and a frictionless comparison does not test whether the network has learned anything beyond what a simpler, already-existing cost-aware adjustment provides. This study tries to close both gaps at once, at the cost of working with a smaller, real, rather than simulated, dataset, drawn from one bounded historical window rather than an arbitrarily long one.

\subsection{Cryptocurrency derivatives markets}

BTC options markets, mainly on Deribit, have grown substantially since their introduction but remain comparatively underexplored in the academic options-hedging literature compared to equity index options such as SPX. We use this market not simply as a convenient data source, but because its data, unlike institutional equity options data, is freely or affordably accessible to an independent researcher, and because its history contains several large, well-documented volatility regimes: the 2020 COVID crash, the 2021 bull market, and the 2022 Terra/Luna and FTX collapses. These give real variation to work with for a regime-conditional comparison of this kind.

\section{Data}

\subsection{Source and acquisition}

We use historical BTC options data from Deribit, obtained through Tardis.dev's freely available historical datasets (Tardis.dev). Tardis.dev distributes full order-book-derived options chain snapshots for the first day of every calendar month without requiring an API key or subscription. Continuous full-history access, comparable to an academic subscription, was quoted at roughly \$650 per month, which was not feasible for this independent, self-funded study. The dataset used here was therefore built from 61 monthly snapshots spanning January 2020 through December 2024, each providing a full day of tick-level options quotes for every actively listed BTC contract.

This acquisition strategy has a direct consequence for how the data has to be modeled. Rather than a continuous multi-year time series, the dataset consists of 61 independent 24-hour windows. We treat each option contract, sampled day pair as an independent hedging episode of at most 24 hourly rebalancing steps, rather than modeling continuous multi-week option lifetimes. This is consistent with the episodic structure used throughout the deep hedging literature, and it does not compromise the within-day analysis central to this study, though it does limit the number of independent observations available for statistical inference, a point we return to in Section~\ref{sec:limitations}.

Raw tick-level data, up to roughly 8.9 million rows for a single day at full order-book granularity, was downsampled to hourly snapshots (the last observation per instrument per hour) to match the hourly rebalancing frequency used throughout this study, and to keep the resulting dataset manageable on standard hardware.

\subsection{Cleaning and filtering}

Three filters were applied to the raw hourly dataset, starting from 892{,}912 rows across 22{,}866 unique instruments.

First, a liquidity filter: rows without both a bid and an ask quote, meaning no two-sided market at that hour, were dropped. This removed 27.0\% of rows. Second, a moneyness filter: contracts with moneyness (strike over spot) outside the range 0.5 to 2.0 were excluded, removing a further tail of illiquid, deep in- or out-of-the-money strikes. Third, a relative spread filter: quotes with a bid-ask spread exceeding 50\% of the mid-price were dropped. This last filter was added after a diagnostic pass showed that Deribit's minimum price tick size produces artificial spikes in relative spread at exact fractional values, such as 0.67 or 1.0, for very low-priced contracts, regardless of true liquidity. Capping relative spread directly removes this artifact rather than relying on the moneyness filter as an imperfect proxy for it.

The final cleaned dataset contains 490{,}229 rows across the full date range. The number of retained rows grows over time, from 138 in 2019 to 189{,}881 in 2024, which reflects the genuine maturation of Deribit's options market rather than any bias introduced by the filtering.

\begin{figure}[h]
\centering
\includegraphics[width=0.95\textwidth]{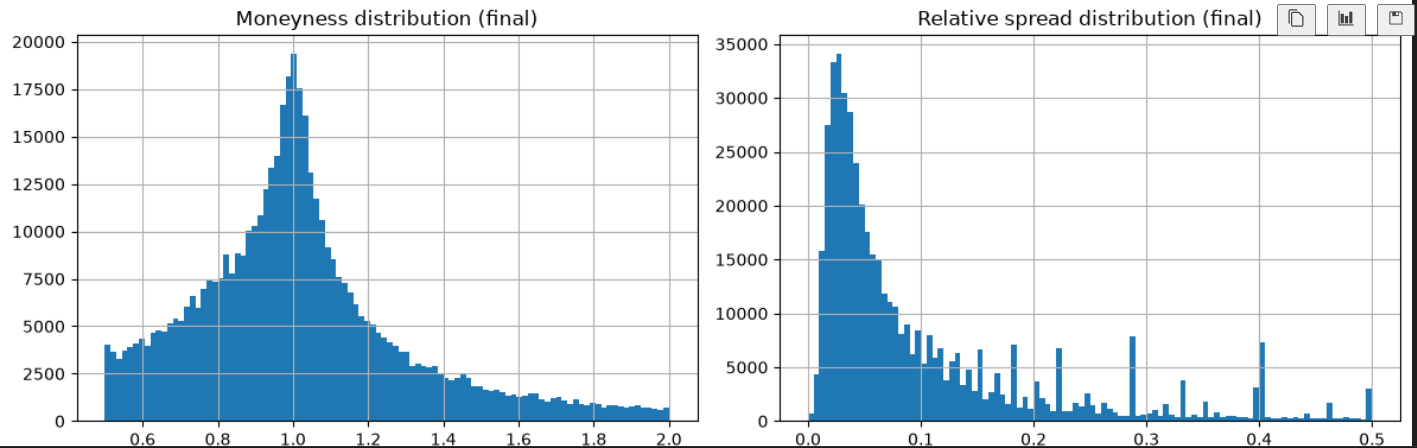}
\caption{Final moneyness and relative bid-ask spread distributions after all cleaning filters. The spike artifacts from minimum tick-size quantization visible in earlier diagnostic passes are no longer present.}
\label{fig:moneyness}
\end{figure}

\subsection{Feature engineering}

For each retained quote we compute time-to-maturity in days and years, moneyness (strike over spot), mid-price (the average of bid and ask, converted from BTC- to USD-denominated using the contemporaneous underlying price), and bid-ask spread, both in absolute terms and relative to the mid-price.

\subsection{Regime classification and train, validation, test split}

Because of the sampled-day structure described above, continuous realized-volatility estimation is not reliable here. Interpolating between monthly snapshots would understate volatility by construction, since linear interpolation has zero realized variance. We instead classify each sampled day's volatility regime using the median at-the-money implied volatility (moneyness between 0.95 and 1.05, using the \texttt{mark\_iv} field) observed on that day. This has the added benefit of using a forward-looking, market-derived volatility measure rather than a backward-looking one, which is arguably more relevant to a study of options hedging in the first place.

\begin{figure}[h]
\centering
\includegraphics[width=0.95\textwidth]{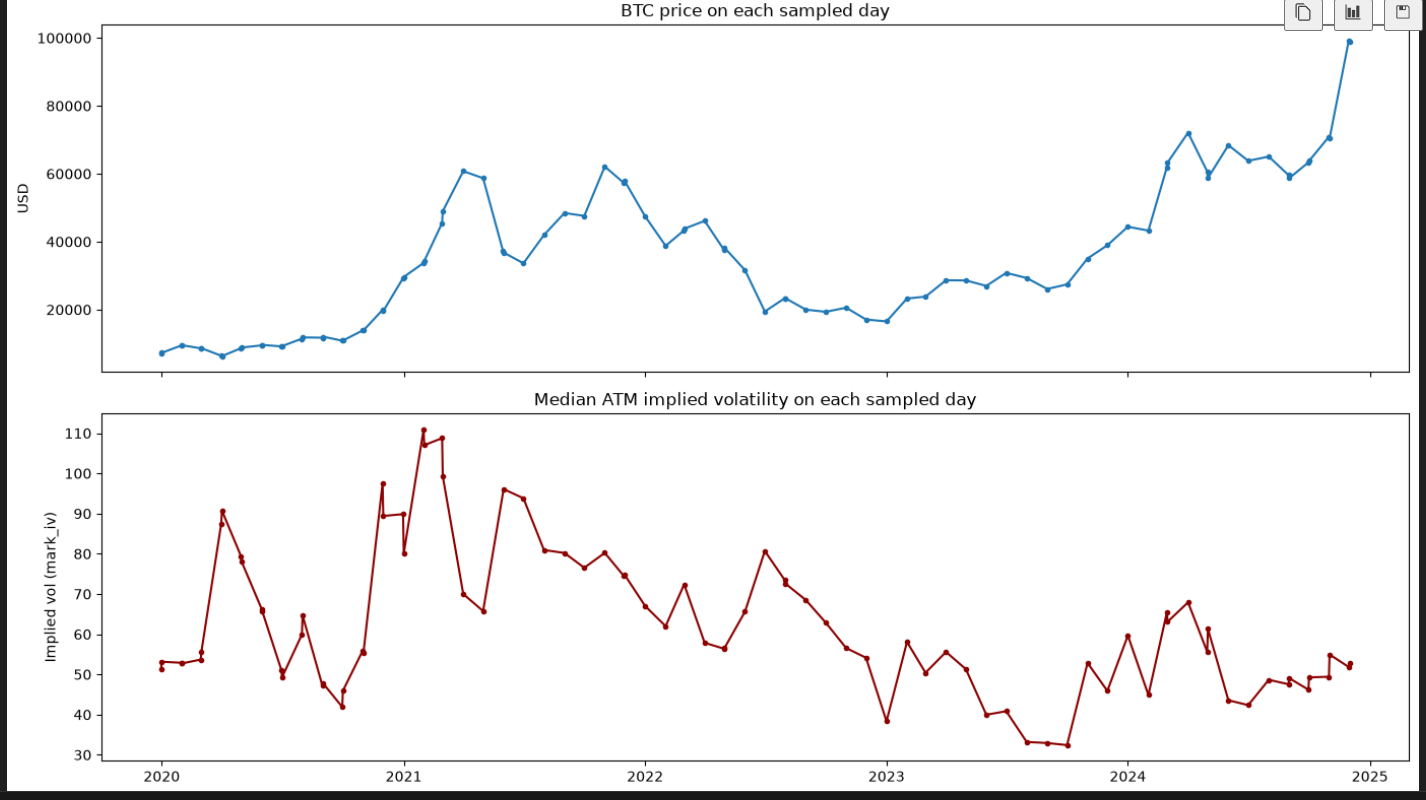}
\caption{BTC price and median at-the-money implied volatility on each of the 61 sampled trading days, used for regime classification and to choose train, validation, and test split boundaries.}
\label{fig:regime}
\end{figure}

Implied volatility across the 61 sampled days ranges from 32\% to 111\%, with a mean of 63\%. It is visibly elevated during the March to April 2020 COVID crash, peaking near 90\%, and during the 2021 bull market, peaking near 111\%, and then declines gradually through 2022 and 2023 before rising again in 2024.

The dataset is split chronologically, with no shuffling, to avoid look-ahead bias:

\begin{itemize}
\item \textbf{Train:} December 2019 through December 2022 (187{,}961 rows). This period captures the COVID crash, the 2021 bull market, and the 2022 Terra/Luna and FTX collapses.
\item \textbf{Validation:} January through August 2023 (64{,}474 rows), a comparatively calm period.
\item \textbf{Test:} September 2023 through December 2024 (237{,}894 rows), which includes a renewed volatility regime tied to the 2024 BTC price rally.
\end{itemize}

We flag here, for clarity ahead of the results in Section~\ref{sec:results}, that the test period spans a single continuous historical window dominated by a strong upward price trend. This shapes both the sign of the P\&L results (Section~\ref{sec:results}) and the scope of the claims this study can support (Section~\ref{sec:limitations}).

\section{Methodology}

\subsection{Problem setup}

We simulate hedging a short position in a single BTC option, taking the perspective of an option seller hedging their exposure, rebalancing hourly using BTC as the hedging instrument. At each step $t$, the option's own mark-to-market P\&L, since the position is short, is the negative of the change in the option's value:
\begin{equation}
\text{P\&L}_{opt,t} = -\left(V_t - V_{t-1}\right), \label{eq:optpnl}
\end{equation}
where $V$ is the option's mid-price in USD. The hedge portfolio's P\&L from holding a BTC position of size $\delta$ established at the previous step is
\begin{equation}
\text{P\&L}_{hedge,t} = \delta_{t-1}\left(S_t - S_{t-1}\right), \label{eq:hedgepnl}
\end{equation}
and the transaction cost of rebalancing from $\delta_{t-1}$ to $\delta_t$ is
\begin{equation}
C_t = \left|\delta_t - \delta_{t-1}\right| S_t \frac{k}{2}. \label{eq:cost}
\end{equation}
The terminal hedging error for an episode of length $T$ is the sum of all three components across every step:
\begin{equation}
\Pi = \sum_{t=1}^{T}\left(\text{P\&L}_{opt,t} + \text{P\&L}_{hedge,t} - C_t\right), \label{eq:terminal}
\end{equation}
and this quantity, $\Pi$, computed once per episode per strategy, is what every metric reported in Section~\ref{sec:results} is derived from.

\subsection{Transaction cost assumption}

Hedging trades in BTC are assumed to incur a round-trip transaction cost rate of $k = 5$ basis points (0.0005), applied as half the round-trip rate to each individual trade, so 2.5 basis points per trade, as shown in Equation~\eqref{eq:cost}. This is broadly consistent with typical BTC spot and perpetual futures spreads on major exchanges. We deliberately avoid using the option's own, often much wider, bid-ask spread as the hedging cost proxy. An earlier version of this pipeline did exactly that and produced an implausible result, with Leland-adjusted volatility exceeding 300\%, more than double the underlying implied volatility, precisely because illiquidity in the options market is not representative of the cost of trading the underlying itself.

\subsection{Classical benchmarks}

\textbf{Black-Scholes (BS) delta.} Computed using Equation~\eqref{eq:d1}, using each contract's own quoted implied volatility (\texttt{mark\_iv}) rather than a separately estimated volatility, which avoids introducing an additional source of model risk. This strategy rebalances to the exact target delta at every hourly step, so 24 out of 24 trades per full episode.

\textbf{Leland's adjusted-volatility hedge.} BS delta computed with the inflated volatility from Equation~\eqref{eq:leland}, using $k = 5$ basis points and $\Delta t$ equal to one hour, expressed in years. This strategy also rebalances every step, but to a somewhat less aggressive target than plain BS delta.

\textbf{Whalley-Wilmott no-trade band.} The band from Equation~\eqref{eq:wwband}, around the BS delta target. The position is held constant whenever it stays within the band $[\Delta - H, \Delta + H]$, and rebalanced only to the nearest band edge, not the exact target, when the band is breached. We settled on $\lambda = 60$ by inspection, picking a value that produces a reasonable number of rebalances per episode. An earlier attempt at calibration, $\lambda = 0.01$, produced a band nearly as wide as delta's entire possible range, which caused the strategy to trade only once per episode in early diagnostic runs.

\begin{figure}[h]
\centering
\includegraphics[width=0.75\textwidth]{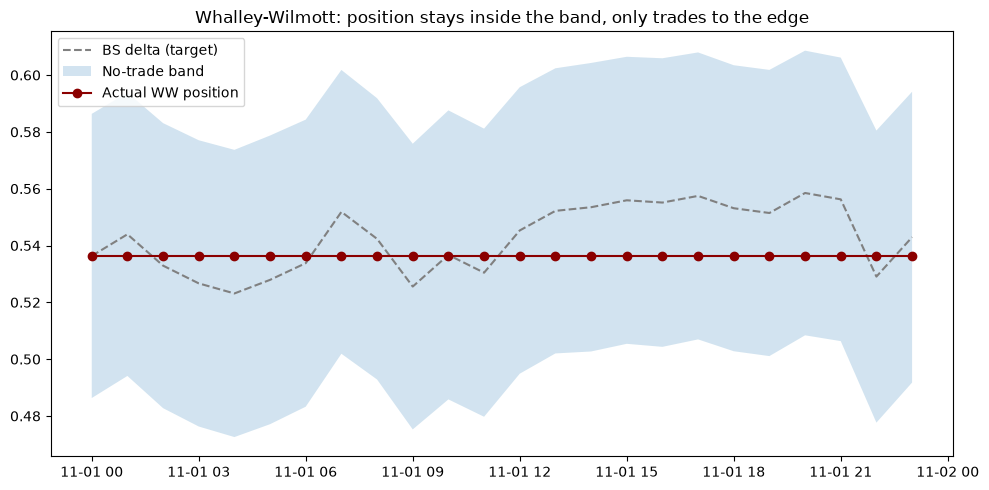}
\caption{Whalley-Wilmott no-trade band (calibrated at $\lambda = 60$) around the Black-Scholes delta target. The resulting position holds flat inside the band and rebalances only to the band edge when the band is breached.}
\label{fig:wwband}
\end{figure}

\subsection{Deep hedging models}

\textbf{State representation.} At each hourly step the policy observes five features: moneyness, time-to-maturity in years, implied volatility (capped at 300\% to limit the influence of a handful of extreme outliers), the contemporaneous BS delta, included as a directly useful reference signal rather than something the network has to rediscover from scratch, and a call or put indicator. Features are standardized to zero mean and unit variance using statistics computed on the training set only.

\textbf{Architectures.} We test two policy architectures. The first is an LSTM (Hochreiter \& Schmidhuber, 1997) with 32 hidden units and a single layer, processing the full episode sequence and able to draw on the entire prior history of the episode through its recurrent hidden state. The second is a feedforward network with two hidden layers of 32 units and ReLU activations, applied independently at each timestep to a fixed 4-step lookback window of features, with no memory beyond that window. Both output a scalar hedge position through $1.5\tanh(\cdot)$, which bounds the output to $[-1.5, 1.5]$, modestly wider than BS delta's $[-1, 1]$ range.

\textbf{Loss function.} The primary training objective is the conditional value-at-risk (CVaR) at confidence level $\alpha = 0.95$ of terminal hedging losses across a training batch of $N$ episodes, following Rockafellar and Uryasev (2000):
\begin{equation}
\text{CVaR}_\alpha = \frac{1}{\lceil (1-\alpha)N \rceil} \sum_{i \,\in\, \text{worst } (1-\alpha)N} L_i, \label{eq:cvar}
\end{equation}
where $L_i = -\Pi_i$ is the loss of episode $i$, and the sum is taken over the worst $(1-\alpha)N$ episodes in the batch. We additionally test an explicit turnover penalty added to this loss, a constant weight $\beta$ multiplying mean per-episode turnover, at two different weights ($\beta = 5$ and $\beta = 100$) across the two architectures, to see whether directly penalizing trading activity is enough to induce sparse, no-trade-band-like behavior.

\textbf{Training.} Episodes are padded to a fixed length of 24 steps with an accompanying mask, and episodes shorter than 4 steps are excluded entirely. Training uses the Adam optimizer (Kingma \& Ba, 2015) with a learning rate of $10^{-3}$ and weight decay of $10^{-5}$, a batch size of 256, and early stopping with a patience of 10 epochs based on validation CVaR loss. The checkpoint with the best validation loss, not the final epoch, is the one retained for evaluation.

\begin{figure}[h]
\centering
\includegraphics[width=0.9\textwidth]{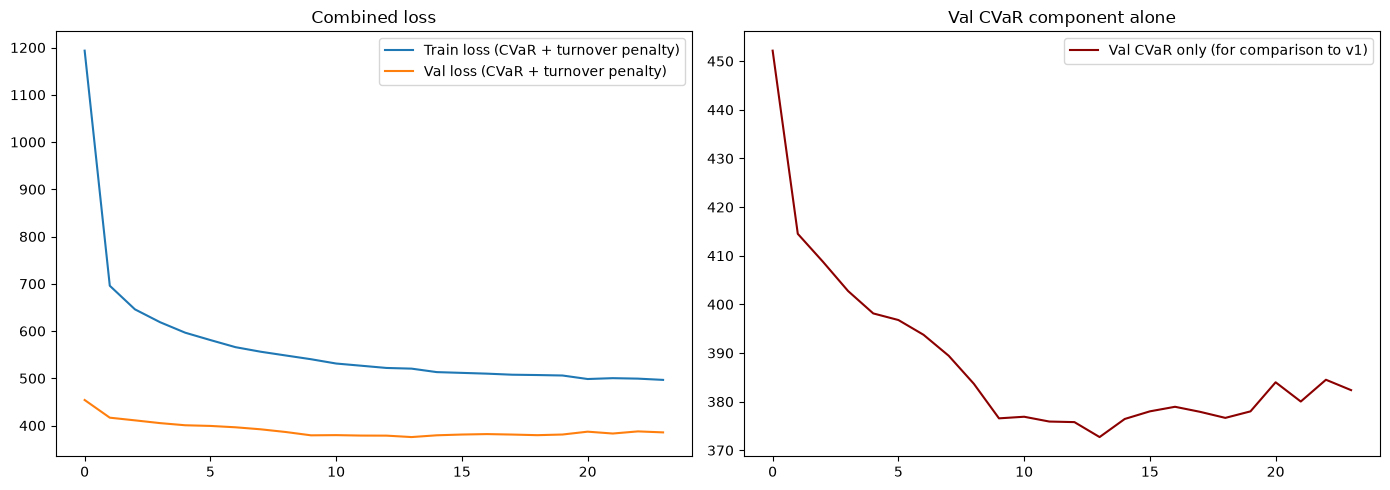}
\caption{Training curves for the turnover-penalized LSTM (v2): combined loss on the left, and the CVaR-only component on the right, showing convergence and the point where early stopping triggers.}
\label{fig:training}
\end{figure}

\subsection{Evaluation metrics}

For each strategy we report mean terminal P\&L (mean of $\Pi$ across episodes), its standard deviation, CVaR at the 95\% level as defined above, mean transaction cost, mean turnover (the sum of absolute position changes), and the mean number of trades per episode. Maximum drawdown is computed by ordering sampled days chronologically, averaging terminal P\&L across episodes within each day, and taking the cumulative sum's peak-to-trough decline. This construction is necessary because episodes are independent by design here, not a single continuous trading path.

\subsection{Statistical testing}

Episodes that share a sampled day also share the same underlying price path and are therefore correlated. Treating them as independent observations would understate the true estimation uncertainty. We instead use a block bootstrap with 5{,}000 resamples, resampling whole sampled days with replacement rather than individual episodes, to build 95\% confidence intervals and approximate two-sided p-values for the difference in mean terminal P\&L, and separately mean cost, between any two strategies.

\section{Results}
\label{sec:results}

\subsection{Overview}

We evaluate six hedging strategies on 11{,}546 out-of-sample test episodes spanning September 2023 through December 2024: Black-Scholes delta, Leland's adjusted-volatility hedge, the Whalley-Wilmott no-trade band, and three deep hedging configurations. These are an LSTM trained with CVaR loss alone (v1), the same LSTM with an added turnover penalty (v2), and a feedforward network with a fixed lookback window trained with a larger turnover penalty. Each episode consists of hourly rebalancing over a single option contract on a single sampled trading day, drawn from 16 unique sampled days. All strategies are evaluated under identical transaction cost assumptions and an identical short-option exposure.

We note before presenting results that this test window includes a sustained, large upward move in BTC price (the 2024 rally). Since every strategy here hedges a short option position, a strong sustained trend works against all of them at once, which is the main reason mean P\&L is negative across every strategy in Table~\ref{tab:summary}. The relevant comparison is therefore relative, which strategy loses the least, and not the sign of any individual number.

\subsection{Summary statistics}

\begin{table}[h]
\centering
\caption{Test-set performance by strategy (11{,}546 episodes).}
\label{tab:summary}
\begin{tabular}{lrrrrr}
\toprule
Strategy & Mean P\&L & CVaR(95\%) & Mean cost & Mean turnover & Mean \# trades \\
\midrule
BS delta & $-38.82$ & $-885.24$ & $9.70$ & $0.668$ & $20.40$ \\
Leland & $-38.91$ & $-886.36$ & $9.68$ & $0.666$ & $20.41$ \\
Whalley-Wilmott & $\mathbf{-30.16}$ & $\mathbf{-864.23}$ & $\mathbf{7.91}$ & $\mathbf{0.545}$ & $\mathbf{2.47}$ \\
Deep hedge (LSTM v1) & $-47.34$ & $-966.23$ & $9.97$ & $0.685$ & $20.44$ \\
Deep hedge (LSTM v2) & $-47.24$ & $-976.87$ & $9.43$ & $0.647$ & $20.44$ \\
Feedforward & $-50.90$ & $-994.85$ & $9.63$ & $0.664$ & $20.44$ \\
\bottomrule
\end{tabular}
\end{table}

Whalley-Wilmott has the best mean P\&L, the best CVaR at the 95\% level, and the lowest cost and turnover of all six strategies, while trading roughly eight times less often than any other approach: 2.47 trades per 24-step episode against roughly 20.4 for everything else. All three neural policy variants underperform all three classical benchmarks on every metric here, and are statistically indistinguishable from one another, which we discuss further in Section~\ref{subsec:significance}.

\subsection{Statistical significance}
\label{subsec:significance}

Because episodes on the same sampled trading day share the same underlying price path, we use a block bootstrap (5{,}000 resamples, blocking by sampled day, giving 16 blocks in the test set) instead of naive per-episode resampling, which would treat correlated episodes as independent and understate the resulting uncertainty.

\begin{table}[h]
\centering
\caption{Block bootstrap results, test set, Whalley-Wilmott versus the classical baselines.}
\label{tab:bootstrap1}
\begin{tabular}{llrrr}
\toprule
Comparison & Metric & Mean diff. & 95\% CI & p-value \\
\midrule
WW vs. BS delta & P\&L & $+8.66$ & $[-3.36, 20.74]$ & $0.164$ \\
WW vs. Leland & P\&L & $+8.75$ & $[-2.95, 20.73]$ & $0.154$ \\
WW vs. BS delta & Cost & $-1.79$ & $[-2.21, -1.39]$ & $<0.0001$ \\
\bottomrule
\end{tabular}
\end{table}

The cost reduction Whalley-Wilmott achieves is statistically significant and economically meaningful, an average saving of \$1.79 per episode against BS delta, driven mechanically by trading roughly 88\% less often. The P\&L and CVaR advantages point the same direction but do not reach significance at this sample size, since the 95\% confidence interval for the P\&L difference crosses zero. We think this comes down to the limited number of independent sampled days available, only 16 in the test set, which caps the effective sample size for the block bootstrap regardless of how many individual option episodes fall within those days.

\begin{table}[h]
\centering
\caption{Block bootstrap results, test set, deep hedging configurations versus the benchmarks.}
\label{tab:bootstrap2}
\begin{tabular}{llrrr}
\toprule
Comparison & Metric & Mean diff. & 95\% CI & p-value \\
\midrule
LSTM v2 vs. WW & P\&L & $-17.08$ & $[-30.39, -4.91]$ & $0.0044$ \\
LSTM v2 vs. WW & Cost & $+1.52$ & $[1.14, 1.92]$ & $<0.0001$ \\
LSTM v2 vs. BS delta & P\&L & $-8.42$ & $[-17.21, 0.02]$ & $0.051$ \\
LSTM v2 vs. LSTM v1 & P\&L & $+0.11$ & $[-9.93, 9.62]$ & $0.964$ \\
Feedforward vs. WW & P\&L & $-20.74$ & $[-33.72, -7.31]$ & $0.0044$ \\
Feedforward vs. LSTM v2 & P\&L & $-3.66$ & $[-14.89, 8.07]$ & $0.506$ \\
\bottomrule
\end{tabular}
\end{table}

All three neural configurations perform significantly worse than Whalley-Wilmott on both P\&L and cost. Against BS delta, LSTM v2's P\&L gap is negative and borderline, with a p-value of 0.051 that narrowly misses the conventional 0.05 threshold. The turnover penalty added in LSTM v2 produced a statistically significant cost reduction relative to v1 (mean difference $-0.54$, $p < 0.0001$), but no matching improvement in P\&L ($p = 0.964$). We trained the feedforward network with a turnover penalty weight twenty times larger than LSTM v2's, specifically to test whether recurrence or penalty strength was the missing ingredient. It performed statistically indistinguishably from the LSTM ($p = 0.506$) and significantly worse than Whalley-Wilmott ($p = 0.0044$), which rules out both explanations.

\subsection{Policy interpretation}

\begin{figure}[h]
\centering
\includegraphics[width=0.85\textwidth]{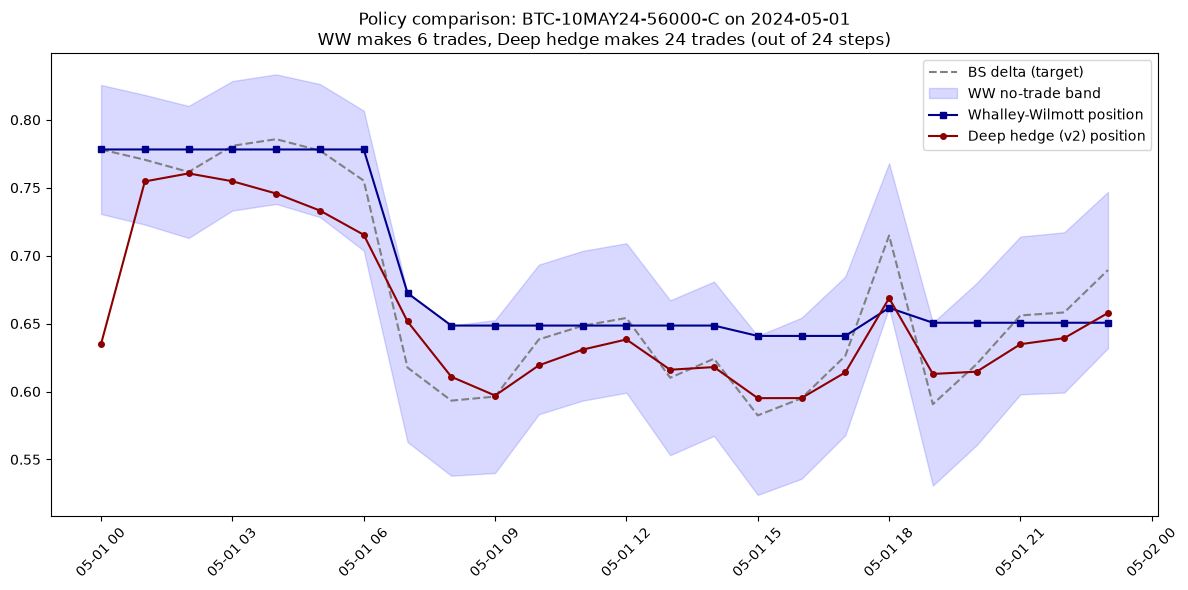}
\caption{Policy comparison on a representative near-the-money test episode. Whalley-Wilmott (blue) holds its position flat inside the no-trade band, executing 6 trades across 24 steps. The deep hedging model (red) tracks the Black-Scholes delta (gray dashed) closely, trading at nearly every step.}
\label{fig:policy}
\end{figure}

Figure~\ref{fig:policy} compares the three policies' hedge positions on a representative near-the-money test episode, BTC-10MAY24-56000-C, from May 1, 2024. Whalley-Wilmott holds its position flat for extended stretches whenever the BS delta stays within its calibrated no-trade band, executing only 6 rebalances across the 24 hourly steps. The deep hedging policy, in contrast, closely tracks the BS delta curve at nearly every step, trading 24 out of 24 times. This pattern holds across the full test set: the mean trade count for all three neural configurations, about 20.44, is statistically indistinguishable from BS delta's 20.40, and far above Whalley-Wilmott's 2.47.

This suggests that the LSTM and feedforward policies, even after we added an explicit turnover penalty to the training loss at weights spanning a twenty-fold range, converged on solutions that closely approximate continuous delta-hedging rather than discovering the discrete no-trade region behind Whalley-Wilmott's cost advantage. The turnover penalty measurably reduced the size of individual trades in LSTM v2, with mean turnover falling from 0.685 to 0.647, but it did not reduce how often the model traded. That explains the significant cost improvement between v1 and v2 alongside the unchanged P\&L and slightly worse CVaR.

\subsection{Validation-set consistency check}

\begin{table}[h]
\centering
\caption{Validation-set summary (January to August 2023, 3{,}520 episodes), shown for reference.}
\label{tab:val}
\begin{tabular}{lrrrr}
\toprule
Strategy & Mean P\&L & CVaR(95\%) & Mean cost & Mean \# trades \\
\midrule
BS delta & $-4.94$ & $-353.55$ & $4.02$ & $18.08$ \\
Leland & $-4.81$ & $-352.87$ & $4.02$ & $18.08$ \\
Whalley-Wilmott & $-5.06$ & $-358.26$ & $\mathbf{3.50}$ & $\mathbf{2.06}$ \\
\bottomrule
\end{tabular}
\end{table}

On the calmer validation period, Whalley-Wilmott's P\&L advantage disappears entirely; it is, if anything, marginally the worst of the three strategies on mean P\&L. Its cost advantage, however, persists. This lines up with the test-set finding that Whalley-Wilmott's biggest advantage shows up specifically during the more volatile, trending regime captured in the test period, which includes the 2024 BTC rally, rather than holding uniformly across all market conditions. We treat this as direct, in-sample evidence that the P\&L result is regime-conditional rather than a universal property of the strategy, a point we expand on in Section~\ref{sec:limitations}.

\begin{figure}[h]
\centering
\includegraphics[width=0.9\textwidth]{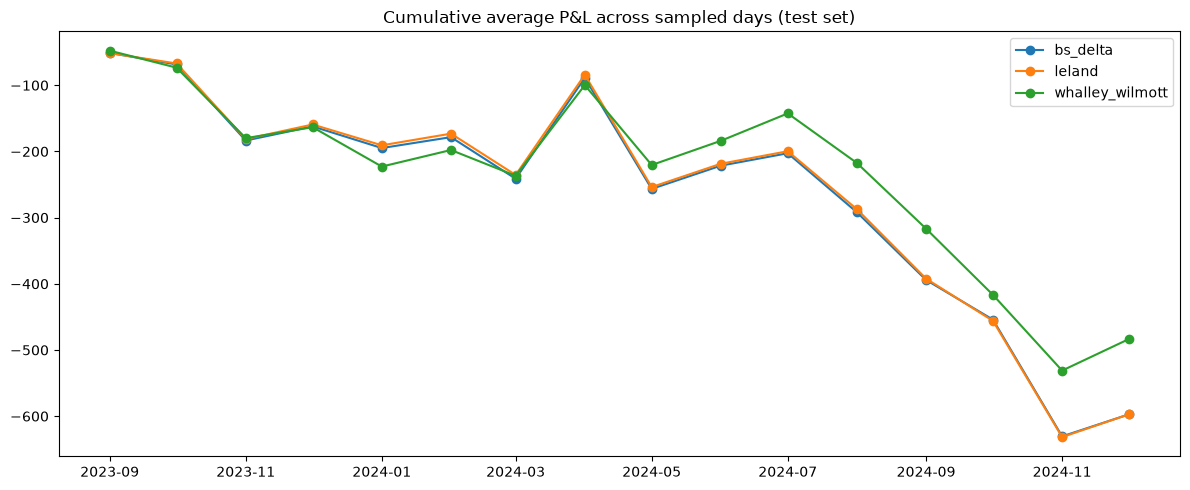}
\caption{Cumulative average terminal P\&L across sampled test-set days, by strategy, used to compute maximum drawdown (Whalley-Wilmott: $-483.38$; BS delta: $-579.38$; Leland: $-580.06$).}
\label{fig:drawdown}
\end{figure}

\section{Discussion}

\subsection{Interpreting the cost result}

The clearest and most statistically solid finding in this study is that Whalley-Wilmott's no-trade band produces a significant, economically meaningful reduction in transaction costs relative to strategies that rebalance continuously, without giving up hedging performance to get it. If anything, its mean P\&L and CVaR are directionally better too, though not significantly so at this sample size. Mechanically this makes sense: a strategy that trades 2.47 times per episode instead of roughly 20.4 will almost always cost less under any positive transaction cost assumption. The fact that this saving did not come at the expense of hedge quality here suggests the calibrated band, with $\lambda = 60$, was reasonably well tuned rather than simply too wide to matter.

\subsection{Why deep hedging underperformed}

All three neural configurations underperformed all three classical benchmarks on every metric we measured, including cost, despite having, in principle, complete freedom to discover a sparse-trading policy similar to Whalley-Wilmott's. We think there are two likely explanations, and they are not mutually exclusive. A robustness check discussed below rules out one plausible alternative explanation, which is architecture choice.

\textbf{Data scale.} The training set contained 8{,}922 episodes of at most 24 timesteps each, several orders of magnitude smaller than the simulated datasets, often $10^5$ to $10^6$ paths, typical in the deep hedging literature. A recurrent or feedforward policy, even one as modest as the ones used here, may simply need more data to reliably discover behaviors like discrete no-trade regions that are not directly encoded in the loss function's gradient at each step.

\textbf{Lack of structural inductive bias toward sparsity.} Whalley-Wilmott's cost efficiency comes from an explicit, hard-coded architectural choice: a conditional decision to trade only when outside the band, rather than a continuous function of state. Neither neural architecture tested here has an equivalent mechanism. We tested this directly by training a feedforward network with a fixed 4-step lookback window and no recurrent memory, using a turnover penalty twenty times larger than LSTM v2's. It performed statistically indistinguishably from the LSTM ($p = 0.506$) and significantly worse than Whalley-Wilmott ($p = 0.0044$). All three neural configurations, the base LSTM, the turnover-penalized LSTM, and the feedforward network, converged to a mean trade count statistically indistinguishable from continuous BS-delta rebalancing, about 20.4 of 24 steps, despite a twenty-fold range in penalty weight.

We think this consistency across architectures and penalty strengths is the most informative negative result in the study. It suggests the gap is not really about a specific network's capacity or a poorly tuned hyperparameter, but about the absence, in all three configurations, of any structural mechanism analogous to Whalley-Wilmott's explicit trade or no-trade rule. A policy architecture that explicitly parameterizes this kind of decision, for instance through a learned threshold or gating mechanism, may be needed to close the gap, and seems like a natural next step for future work.

\subsection{Limitations}
\label{sec:limitations}

\textbf{Scope of the time period tested.} The headline result in this study comes from a single, continuous out-of-sample test window (September 2023 to December 2024), which happens to include a large, sustained BTC price rally. This is a genuine limitation on how far the finding can be generalized in time, and we want to be explicit about it rather than let it be inferred. It is not accurate to describe this study as showing that deep hedging ``does not work'' in general; the more precise and defensible claim is that, in this specific window, under this specific cost assumption, it did not outperform the classical benchmarks, and the comparative advantage of Whalley-Wilmott specifically appears tied to the volatile, trending character of that window. The validation-period check in Section~\ref{sec:results} supports this directly: in a calmer, non-trending period, Whalley-Wilmott's P\&L edge disappears while its cost edge persists, meaning the two results (cost savings, P\&L improvement) do not have the same evidentiary strength or the same scope. A study spanning multiple independent test windows, each covering a different regime type in isolation, is the natural way to test this more rigorously than we have here, and is the most direct way future work could either confirm or overturn the pattern we report.

\textbf{Historical data coverage.} This study relies on Deribit's freely available first-of-month historical options data, which gives one real trading day per calendar month rather than continuous daily coverage. This shapes the analysis in two specific ways. First, the test set contains only 16 unique sampled days, which limits the effective sample size for the block bootstrap regardless of how many individual option contracts fall within those days, and this directly explains why the cost result, a large and consistent effect, reached significance while the noisier P\&L and CVaR results did not. Second, within-month dynamics are simply not observed, since each sampled day is treated as an independent 24-hour episode.

\textbf{Transaction cost model.} We use a fixed 5 basis point round-trip cost applied uniformly across all time periods, moneyness levels, and volatility regimes, rather than a time-varying or liquidity-conditional cost model.

\textbf{Risk-aversion parameter calibration.} The Whalley-Wilmott band width depends on a risk-aversion parameter, $\lambda$, that has no directly observable market value. We calibrated $\lambda = 60$ by inspection rather than through an independent estimation procedure. A formal sensitivity analysis across a range of $\lambda$ values would be a reasonable robustness check for future work.

\textbf{Deep hedging architecture scope.} We tested three neural configurations across two architectures and two loss formulations, which is more thorough than what was originally planned, and the consistency of the result across all three strengthens our confidence that it is not just an artifact of one particular implementation choice. It remains possible that a fundamentally different approach, such as an architecture with an explicit gating mechanism, or training on a much larger dataset, would close some or all of the gap. This study shows that several reasonable, fairly standard implementations do not close it, not that no implementation could.

\textbf{Single asset class.} All results here are specific to BTC options on Deribit. Given the known differences in market structure, participant composition, and liquidity between crypto derivatives and traditional equity index options, whether these findings generalize to other underlyings is an open question rather than something we can assume.

\section{Conclusion}

This study set out to answer a fairly narrow question: under realistic transaction costs, and compared fairly against classical hedging methods that were themselves built to handle those costs, does a standard deep hedging policy actually outperform them? Using five years of real BTC options data, we find that, within the specific test window studied, it does not. The Whalley-Wilmott no-trade band, a closed-form strategy that is decades old at this point, achieves significantly lower transaction costs and directionally better risk-adjusted returns than every deep hedging configuration we tested, including after a genuine attempt to close the gap with an explicit turnover penalty, tried at two weights across two architectures. We are careful to state this as a time-bounded, regime-linked finding rather than a universal one, since the validation-period check in Section~\ref{sec:results} shows the P\&L advantage specifically weakening in a calmer market.

We think the more useful contribution of this study is the mechanistic explanation behind that result, not just the result itself. All three neural configurations converged on a policy that closely approximates continuous Black-Scholes delta-hedging, trading at nearly every timestep no matter the penalty weight or the architecture. This suggests that the advantage of a method like Whalley-Wilmott's does not come simply from being aware of transaction costs, since all of our neural policies had that awareness built into their loss function, but from an explicit structural mechanism, a hard trade or no-trade decision, that a standard continuous-output policy trained only to minimize a risk-adjusted loss did not discover on its own, at least not at the data scale available here.

\textbf{Future work.} Four directions follow fairly directly from the limitations discussed in Section~\ref{sec:limitations}. The first, and most direct response to the temporal scope limitation, is repeating this comparison across multiple independent test windows spanning different regime types in isolation (a purely calm period, a purely crash period, a purely trending period), to determine how much of the reported advantage is regime-specific rather than general. The second is training on a substantially larger dataset, likely requiring either paid continuous historical data access or a carefully validated simulation-based pretraining stage, to see whether the neural policies' underperformance is a data-scale artifact rather than something more fundamental. The third is testing architectures with an explicit gating or thresholding mechanism, closer in spirit to the classical no-trade band, to see whether structural inductive bias, rather than penalty shaping alone, is what is actually needed to induce sparse trading. The fourth is extending the comparison to other underlyings, such as equity index options, where institutional data access allows for continuous historical coverage, to see whether the findings here are specific to BTC options market structure or hold more broadly.


\begin{thebibliography}{99}

\bibitem[Black and Scholes(1973)]{black1973pricing}
Black, F., \& Scholes, M. (1973). The pricing of options and corporate liabilities. \textit{Journal of Political Economy}, 81(3), 637--654.

\bibitem[Buehler et al.(2019)]{buehler2019deep}
Buehler, H., Gonon, L., Teichmann, J., \& Wood, B. (2019). Deep hedging. \textit{Quantitative Finance}, 19(8), 1271--1291.

\bibitem[Cao et al.(2020)]{cao2020deep}
Cao, J., Chen, J., Hull, J., \& Poulos, Z. (2020). Deep hedging of derivatives using reinforcement learning. \textit{Journal of Financial Data Science}, 3(1), 10--27.

\bibitem[Carbonneau(2021)]{carbonneau2021deep}
Carbonneau, A. (2021). Deep hedging of long-term financial derivatives. \textit{Insurance: Mathematics and Economics}, 99, 327--340.

\bibitem[Hochreiter and Schmidhuber(1997)]{hochreiter1997long}
Hochreiter, S., \& Schmidhuber, J. (1997). Long short-term memory. \textit{Neural Computation}, 9(8), 1735--1780.

\bibitem[Kingma and Ba(2015)]{kingma2015adam}
Kingma, D. P., \& Ba, J. (2015). Adam: A method for stochastic optimization. \textit{Proceedings of the 3rd International Conference on Learning Representations (ICLR)}.

\bibitem[Leland(1985)]{leland1985option}
Leland, H. E. (1985). Option pricing and replication with transaction costs. \textit{Journal of Finance}, 40(5), 1283--1301.

\bibitem[Rockafellar and Uryasev(2000)]{rockafellar2000optimization}
Rockafellar, R. T., \& Uryasev, S. (2000). Optimization of conditional value-at-risk. \textit{Journal of Risk}, 2(3), 21--41.

\bibitem[Tardis.dev]{tardisdev}
Tardis.dev. Historical tick-level market data for cryptocurrency exchanges. \url{https://tardis.dev}

\bibitem[Whalley and Wilmott(1997)]{whalley1997asymptotic}
Whalley, A. E., \& Wilmott, P. (1997). An asymptotic analysis of an optimal hedging model for option pricing with transaction costs. \textit{Mathematical Finance}, 7(3), 307--324.

\end{thebibliography}
\end{document}